# Spin dynamics, short-range order and superparamagnetism in superconducting ferromagnet $RuSr_2Gd_{1.4}Ce_{0.6}Cu_2O_{10-\delta}$

Anuj Kumar[1,2], R. P. Tandon[2] and V. P. S. Awana[1,*]

[1]*Quantum Phenomena & Application Division, National Physical Laboratory (CSIR) Dr. K. S. Krishnan Road, New Delhi-110012, India*

[2]*Department of Physics and Astrophysics, University of Delhi, North Campus New Delhi-110007, India*

# Abstract

We report structural, detailed DC and linear/non-linear AC, isothermal and thermoremanent magnetization study of the rutheno-cuprate superconducting ferromagnet $RuSr_2Gd_{1.4}Ce_{0.6}Cu_2O_{10-\delta}$ (GdRu-1222). Structural analysis, by employing Rietveld refinement of X-ray diffraction pattern, reveals that GdRu-1222 crystallizes in tetragonal phase with *I4/mmm* space group. GdRu-1222 is a reported superconducting ferromagnet with Ru spins magnetic ordering at around 110 K and superconductivity below 40 K in $Cu-O_2$ planes. Detailed linear/non-linear first and higher order harmonic of AC susceptibility studies unveiled the complex magnetism of GdRu-1222. A frequency dependent cusp is observed in AC susceptibility ($\chi_{ac}$) vs. *T* measurements. The change in cusp position with applied frequency followed the well known *Vogel-Fulcher law*, which is a feature to describe a spin-glass (SG) system with possibility of embedded homogeneous/non-homogeneous magnetically interacting/non-interacting ferromagnetic clusters. Such an interpretation is also supported by thermoremanent magnetization (TRM) study at *T* = 60 K. Detailed interpretation of AC magnetization results revealed the formation of magnetic (ferromagnetic) homogenous/non-homogenous clusters of different sizes embedded in spin-glass (SG) matrix. The magnetization vs. applied field loops do not saturate, even at high applied fields (50 kOe), resulting in the short-range magnetic ordering in the system, which causes the formation of clusters that freeze at low temperatures. Temperature variation of first- and third-order susceptibility harmonics show good agreement with Wohlfarth's model (WM), leading to the superparamagnetism (SPM) state. Detailed magnetization (DC and AC both) results and their analysis helped in explaining the temperature dependent magnetism of the GdRu-1222 system.

**PACS Nos.** *74.70 Pq, 75.30.Cr, 75.78.-n & 75.50.Vv*

**Keywords:** *Ruthenates, Susceptibility, Magnetization dynamics & Superparamagnetism*

# I. Introduction

Discovery of co-existing superconductivity (SC) and weak ferromagnetic order (W-FM) in the hybrid rutheno-cuperate systems $RuSr_2(Eu,Gd,Sm)_{1.5}Ce_{0.5}Cu_2O_{10-\delta}$ (Ru-1222) and $RuSr_2(Eu,Gd,Sm)Cu_2O_{8-\delta}$ (Ru-1212) are particularly interesting because the magnetic ordering temperature or Curie temperature ($T_C$) is much higher than the superconducting transition temperature ($T_c$) [1-5]. Despite extensive research on these materials, some unanswered questions remain unsolved. In particular, the possibility of magnetic ordering of the weak ferromagnetic (W-FM) [1] or ferromagnetic type [6], as originally reported based on bulk magnetization measurements, generated additional excitement with some agnosticism. The reason behind that the dipolar and exchange fields generated by a FM or W-FM $Ru-O_2$ layer in close vicinity to the $Cu-O_2$ layers could act as pair breakers or stop singlet-pair formation altogether. Density functional theory [7] concluded some of these concerns by showing that these dipolar and exchange fields are weak enough in Ru-1222 and hence singlet pairing can still survive in the $Cu-O_2$ layers with a modulated superconducting (SC) order parameter. This depends on whether the Ru magnetization is parallel or perpendicular to the $Ru-O_2$ layers. One of the most controversial questions is the exact type of magnetic ordering in Ru-1222 family. Both Ru-1212 and Ru-1222 possess two $Cu-O_2$ planes and one $Ru-O_2$ layer in a tetragonal unit cell with space group *P4/mmm* and *I4/mmm* respectively. The structure of Ru-1212 is related to the structure of well known $CuBa_2YCu_2O_{7-\delta}$ (Cu-1212) such that the Ba ion is replaced by Sr ion and $Cu-O_{1-\delta}$ chain is replaced by $RuO_{2-\delta}$ sheet. While in case of Ru-1222 a fluorite type block $(R,Ce)O_{2-\delta}$ (R = Eu & Gd) is inserted between two Ru-1212 unit cells and each unit cell of Ru-1212 is shifted by (*a*/2, *a*/2) coordinate positions [8, 9]. Studies of DC magnetization [10], muon-spin rotation (μSR) [11], as well as Mossbauer spectroscopy [12] indicated a double magnetic transition, which has been explained as due to the presence of some Ru-1212 impurity phase in Ru-1222. Further, granularity and clustering in the system had been considered to lead to various phenomena with the contradictory explanations. Phase separation into ferromagnetic (FM) clusters and paramagnetic matrix followed by an antiferromagnetic (AFM) transition has been assumed to establish long-range order in the system [13]. In contrast to this, neutron diffraction data [14, 15] did not show any long-range order, which is further contradicted by Mclaughlin *et al.* [16] who observed clear magnetic scattering in their neutron-diffraction measurements on $RuSr_2Y_{1.5}Ce_{0.5}Cu_2O_{10-\delta}$ (YRu-1222), indicating antiferromagnetic alignment of Ru spin (Ru moments along the *c*-axis). On the other hand the proposed magnetically frustrated spin-glass (SG) [17-19] opposed the claims of long-range magnetic order in the system. The neutron diffraction data could not be modeled with a simple G-type AFM structure and arguments were put in favor of both FM and AFM Ru-Ru coupling being simultaneously present along the *c*-axis [16]. Neutron diffraction study also failed to observe a net FM component in Ru-1222 compounds, with an upper limit of ~ 3$\mu_B$/Ru as calculated from magnetization measurements. Recently, it was proposed, in $Nb_{1-x}Ru_x$-1222 system, that there are interacting clusters at the Ru-$O_2$ planes without any long-range magnetic order [20]. On the other hand, the slow spin dynamics [13] suggested that the FM clusters in Ru-1222 could exhibit superparamagnetism (SPM). Whereas, observation of the frequency dependent peak shift of the AC susceptibility as a function of temperature along with thermoremanent magnetization (TRM) measurements [21] have indicated spin glass (SG) behavior in $RuSr_2Gd_{1.5}Ce_{0.5}Cu_2O_{10-\delta}$, which also contradicts the existence of long-range order in the system. The magnetic behavior in Ru-1222 is more challenging and complex as compared to Ru-1212. One of the drawbacks with the rutheno-cuperate based superconductors is that the most effective technique for studying magnetism i.e., neutron diffraction, is not suitable. This is because both Ru-1212 and Ru-1222 systems forms with only Eu, Gd and Sm, which are high neutron absorbers. Hence, the non-linear susceptibility is the most effective tool to investigate the complex magnetic behavior of Eu, Gd, and Sm based rutheno-cuprates. In particular, strong evidences of spin-glass (SG) behavior was observed in $Gd_{1.5}Ce_{0.5}$Ru-1222 [21].

Superparamagnetism is nothing but an ensemble of nano particles, in which the inter-particle magnetic interactions are sufficiently weak. When the inter-particle interactions are non-negligible, the system eventually shows collective behavior, which overcomes the individual anisotropy properties of the particles. On the other hand at sufficiently strong interactions a magnetic nanoparticle ensemble can exhibit super spin-glass (SSG), which is similar to the spin-glass (SG) systems in bulk materials. In the typical superparamagnetic state, below the temperature called the blocking temperature ($T_B$), the anisotropy energy is greater than the thermal energy so that easy axis of magnetization in clusters orient in same direction. The spin-glass (SG) state [22, 23] is a low temperature phenomenon that occurs due to the disorder and frustration in the system. Frustration is created due to the competing ferromagnetic and antiferromagnetic interactions between the neighboring spins. At a particular temperature, called freezing temperature ($T_f$), all spins freeze in a random direction, in order to minimize the total energy of the system. Non-linear AC susceptibility is a very effective tool to investigate the spin-glass (SG) and superparamagnetism (SPM) states in the material. Because the measurements can be performed at very low applied AC fields, hence any small change in the magnetic susceptibility due to the phase transition can be observed. Otherwise these small changes in AC magnetic susceptibility could have been masked by the high applied field. Non-linear complex AC susceptibility in presence of an excitation field $H_{ac}$ can be interpreted as:

$$M = M_0 + \chi_1 H_{ac} + \chi_2 H_{ac}^2 + \chi_3 H_{ac}^3 + ..., \qquad (1)$$

where $\chi_1$, $\chi_2$ and $\chi_3$ are the first, second and third-order harmonic susceptibilities respectively [22, 23].

In this paper we extend our investigation [18] of the complex magnetic behavior in rutheno-cuprate systems. The temperature dependent DC, linear/non-linear AC magnetization (frequency and field dependence) and thermoremanant magnetization (TRM) of $RuSr_2Gd_{1.4}Ce_{0.6}Cu_2O_{10-\delta}$ (GdRu-1222) sample are studied in detail to understand the spin-glass (SG) with ferromagnetic clusters (FM) and superparamagnetism (SPM) states. Specially first and third-order harmonic of AC susceptibility are discussed in detail to probe the superparamagnetism (SPM) state in this compound. A temperature dependent scenario of the complex magnetism of GdRu-1222 is presented.

## II. Experimental details

Polycrystalline bulk sample of $RuSr_2Gd_{1.4}Ce_{0.6}Cu_2O_{10-\delta}$ (GdRu-1222) was synthesized through solid state reaction route from stoichiometric powders of purity 99.9% $RuO_2$, $SrCO_3$, $Gd_2O_3$, $CeO_2$ and $CuO$. These mixtures were ground together in an agate and calcined in air at $1020^oC$, $1040^oC$ and $1060^oC$ each for 24 hrs with intermediate grindings. The pressed bar shaped pellet of the sample was annealed in Oxygen atmosphere at $850^oC$, $650^oC$ and $450^oC$ each for 24 hrs, and subsequently cooled down slowly over a span of 12 hrs to the room temperature. X-ray diffraction (XRD) was performed at room temperature in the scattering angular ($2\theta$) range of $20^o$-$80^o$ in equal $2\theta$ step of $0.02^o$ using *Rigaku Diffractometer* with *Cu $K_\alpha$* ($\lambda$ = 1.54Å) radiation. Rietveld analysis was performed using the standard *FullProf* program. Sample is crystallized in tetragonal structure with *I4/mmm* space group. Detailed DC and AC (linear and non-linear) magnetization were performed on Physical Property Measurements System (PPMS-14T, Quantum Design-USA) as a function of both temperature and applied magnetic field. Linear and non-linear AC susceptibilities as a function of temperature (i) in the frequency range 33-9999 Hz and, (ii) in the AC drive magnetic field amplitude variation 1-17 Oe, with zero external DC magnetic fields were also measured on PPMS-14T. Resistivity measurement was performed in zero magnetic fields on a close cycle refrigerator (CCR), in temperature range 12-300 K, designed by Advanced Research System (ARS) USA.

## III. Results and Discussion

Phase purity of complex Rutheno-cuperates GdRu-1222 is very important for a meaningful scientific discussion, because impurities like $SrRuO_3$ (SRO) and $Sr_2RuGdO_6$ ($211O_6$) phase tend to form readily in the host GdRu-1222 matrix. Small impurity of these compounds can alter the net outcome magnetization of GdRu-1222. It is clear that main peaks corresponding to SRO and $211O_6$ phases are not observed within the XRD limit. Observed (open circle) and fitted (solid lines) X-ray patterns for the studied compound $RuSr_2Gd_{1.4}Ce_{0.6}Cu_2O_{10-\delta}$ (GdRu-1222) are shown in figure 1. The structural analysis was performed using the Rietveld refinement analysis by employing the *FullProf* Program. The Rietveld analysis confirms a single phase formation in tetragonal structure with space group *I4/mmm*. All structural parameters (lattice parameters, atomic coordinates and site occupancy), are shown in the Table I.

Typical magnetization curve as a function of temperature under small applied magnetic field of 20 Oe is shown in figure 2. Clearly there is a ferromagnetic like (FM-like) transition observed at around 110 K. We define $T_C$ = 110 K, i.e., Curie temperature corresponding to significant zero-field-cooled (ZFC) and field-cooled (FC) branching of the magnetization curve. The sharp rise of both the ZFC and the FC curves for GdRu-1222 at 110 K shows a PM to FM transition. MT curve exhibits the strong irreversibility between the ZFC and FC, which is typical of a spin-glass (SG) and superparamagnetism (SPM) relaxation phenomena. The ZFC curve of the compound GdRu-1222 has a peak at $T_{cusp}$ = 90 K, while the FC curve is increasing with decrease in temperature. This steady increase of the FC branch at low temperature is interpreted as being caused by the high moment paramagnetic response of the Gd ions. Further on lowering the temperature the GdRu-1222 shows the superconducting transition at $T_{c(dia.)}$ = 27 K. In our previous paper [18] we have shown that EuRu-1222 undergoes the spin-glass transition, with ferromagnetic clusters, at around the peak temperature of $\chi'_1(T)$, (see figure 5(a), ref. 18) resulting in the freezing of spins. Hence the peak temperature is called the freezing temperature ($T_f$). This peak temperature corresponds to the peak in the ZFC curve with a small change in temperature because of the difference in the response of the system to DC and AC fields. Inset of figure 2 shows the *R* vs. *T* behavior of studied GdRu-1222 in zero magnetic fields. Superconducting transition temperature ($T_c$) is seen at around 30 K. As we move from higher to lower temperatures, the resistance increases continuously, exhibiting an insulating type behavior. This insulating behavior continues until the temperature reaches the onset of the superconducting transition temperature. Interestingly, the slope of the resistance curve changes at around 90 K and resistance increases faster as compared to from 300 to 90 K range. In GdRu-1222, the fast increase of resistance below $T_f$ is in contrast to the reported canonical spin-glasses such as AuCr and CuMn, [23], which exhibits a resistivity maximum at $T_f$. These systems are metallic spin glasses, with diluted magnetic impurities, dominated by Ruderman-Kittel-Kasuya-Yoshida (*RKKY*) interaction between impurity spins [24]. It seems the studied GdRu-1222 is not a pure/clean canonical spin-glass system.

To further elucidate upon the magnetic properties of the GdRu-1222 magneto-superconductor, we show isothermal magnetization for various values of applied fields. Figure 3 shows the typical magnetization loops of studied GdRu-1222 sample measured at various temperatures (5, 20, 50, 75, 100, 125, 150 and 200 K) up to the range of ±50 kOe. The isothermal magnetization as a function of applied magnetic field may be expressed as the sum of a linear part and a non-linear part. Means, *M(H) = χH + $σ_s$(H)*. Here the linear contribution *χH* arises from the combined effects of the antiferromagnetic component of Ru spins and the patramagnetic Gd spins. While $σ_s$*(H)* represents the ferromagnetic component of the Ru spin moments. *M(H)* curve below the Curie temperature ($T_C$ = 110 K) shows the FM-like hysteresis behavior with non-saturating magnetization at high fields (up to 50 kOe). This behavior is also observed in the itinerant ferromagnets [25]. This suggests the formation of short range ordered clusters with ferromagnetic coupling between them. At low temperature (say 5 K) the *M(H)* loop has S-type shape, which is the feature of spin-glass (SG) system. In low field range, there is a small opening near the origin which confirms the ferromagnetic nature below the Curie temperature. Representative *M(H)* plots in low field regime (up to 2 kOe) at 5, 20 and 50 K are shown in inset of figure 3. Remember the compound is superconducting below ~ 30 K, hence the *M(H)* at 5 and 20 K is combination of both the diamagnetic (superconductivity) and Ru spins complex magnetic ordering (SG/FM-AFM etc.). As the temperature increase the S-type shape transformed into the linear paramagnetic (PM) above 150 K. At 200 K this shape is completely transformed into the straight line PM state. Both, the presence of ferromagnetic like nature at low temperature range and the absence of saturation magnetization at high field (±50 kOe) are the possible characteristics of a spin-glass (SG) system [26, 27] with possibility of embedded non-interacting homogenous/non-homogenous ferromagnetic clusters. More details in this respect will be provided in next sections. The ferromagnetic behavior at temperature around Curie temperature ($T_C$) can be investigated using $M^2$ vs. $H/M$ (Arrot plots) [28]. The $M^2$ vs $H/M$ curves reveal a linear behavior around $T_C$ and are linear and passing through the origin at $T = T_C$. Additionally, according to the criterion proposed by Banerjee [29], the order of magnetic transition can be determined from the slope of these straight lines. The positive slope corresponds to the second-order-transition, while the negative one corresponds to the first-order-transition. Figure 4 shows the Arrot plots for GdRu-1222 sample in the temperature range of 5 - 125 K. Clearly, in present case the positive slope of $M^2$ vs $H/M$ curves indicate the ferromagnetic phase transition to be of the second order. However, all the curves in Arrot plots are non-linear with a curvature towards the $M^2$-axis without any intercept on the same axis. The curve corresponding to 100 K is passing through the origin, hence this is taken as the Curie temperature of the system, which is in agreement with $T_C = 110$ K being observed from MT (figure 2). It is also seen that curves in Arrot plots do not possess any spontaneous magnetization, which indicates towards short-range ordering in the system [30]. Absence of spontaneous magnetization and existence of short-range ordering confirm the possible indication of spin-glass phase with ferromagnetic clusters.

Further, the existence of spin-glass (SG) state has been confirmed through the time-dependent DC magnetic behavior of the sample. The time response of DC magnetization is very important to reveal the spin dynamics of a spin-glass (SG) system [22, 23]. The remanence magnetization can be measured in two different ways one is the isothermal remanence magnetization (IRM) and other is thermoremanence magnetization (TRM). But here we performed the TRM for the studied GdRu-1222 sample. It was observed that the remanent magnetization is dependent on magnetic history, with IRM(H, T)≤TRM(H, T) for small fields range, while in presence of high enough fields both IRM and TRM saturate at the same value. The field dependence of the IRM and TRM data is reported for classical spin glasses, such as $Eu_{0.3}Sr_{0.7}S$ [31]. We measured TRM for different waiting times ($t_w$ = 100 s and 1000 s) at 60 K (shown in figure 5). The sample was field-cooled (FC) in the presence of 5 kOe field from 200 K to 60 K and after certain waiting time ($t_w$ = 100 s and 1000 s) the applied field was reduced to zero and the corresponding decay of the magnetization was recorded as a function of waiting time. The results show that there is a pronounced waiting time dependence of the relaxation. Various functional forms have been purposed to describe the magnetization as a function of elapsed time and waiting time. One of the most popular relations is the stretched exponential relation,

$$M(t) = M_0[-(t/t_p)^{1-n}]........ (4)$$

where $M_0$ and $t_p$ depend on $T$ and $t_w$, while the number $n$ is only the function of $T$ [32]. The characteristics behavior of $M_{TRM}$ (t) is called the aging effect, which is already reported in other spin-glass (SG) systems [33, 34]. According to the droplet scaling theory the aging affect is ascribed to the thermal growth of spin-glass ordered domains [35]. The observed behavior (figure 5) of TRM for GdRu-1222 is the same as reported for other spin-glass systems [35, 36]. Hence the studied GdRu-1222 is in spin-glass (SG) state with ferromagnetic clusters. The situation will be clear, while we will discuss the AC susceptibility in next sections.

The AC susceptibility ($\chi_{ac}$) technique (with frequency and amplitude variation) is a powerful tool to study the spin-glass (SG) system with ferromagnetic clusters. In the case of spin-glass, both components χ' and χ'' of $\chi_{ac}$, show a frequency dependent cusp. The position of cusp in χ' defines the freezing temperature $T_f$, which is coincident with the temperature of the inflection point in χ''. In order to understand the spin-glass (SG)/cluster-glass (CG) behavior of the compound, we studied its dynamics by AC susceptibility measurements. In figure 6(a) and 6(b) we present the temperature variation of the real ($\chi'_1$) and imaginary ($\chi''_1$) part of the first harmonic of AC susceptibility ($\chi_{ac}$) under different frequencies (ranging 33 to 9999 Hz) of the alternating field. It is observed that before the main PM to FM or spin-glass (SG) transitions, there is a hump seen at around 127 K, called Neel temperature ($T_N$), which is consistent with earlier report on this system [37]. The sharpness of the peak observed in AC susceptibility is an indication of the homogenous phase transition [23]. A paramagnetic (PM) to ferromagnetic (FM) transition, called Curie temperature also observed in both real ($\chi'_1$) and imaginary part ($\chi''_1$) of the AC susceptibility curves. The real ($\chi'_1$) curves exhibit clear peaks while the corresponding imaginary ($\chi''_1$) curves show inflection around the spin-glass (SG) transition temperature or freezing temperature $T_f$ (f). On further lowering the temperature the studied GdRu-1222 shows the clear superconducting transition [$T_{c(sup.)}$] at 30 K and $T_{c(dia.)}$ = 20 K. (shown in figure 6(a)). Inset of figure 6(a) shows that the height of the peak corresponding to the freezing temperature $T_f$ decreases and its position shifts towards the higher temperature with the increasing applied frequency of the alternating field. On the other hand, for imaginary part ($\chi''_1$) the height of the peak increases and also the peak shifts towards the higher temperature (see figure 6(b)). In both cases peak shifts towards the higher temperature but the exact shift is larger for imaginary part($\chi_1''$) than the real one ($\chi_1'$). It is observed that there is a change in freezing temperature with applied frequency of the alternating field. The change in freezing temperature $T_f(\chi')$ ($T_f$ = 92.91 K at $f$ = 33 Hz and $T_f$ = 93.52 K at 9999 Hz) with applied frequency is the characteristics of spin-glass (SG) behavior. The primary criterion to predict the spin-glass (SG) state for a material is $\delta T_f = \Delta T_f/T_f \Delta(\log_{10} f)$, where $\Delta$ represents the change in the corresponding quantity. For spin-glass systems $\delta T_f$ varies in the range of 0.004-0.018, while for superparamagnetism systems it is of the order of 0.3-0.5 [22]. The initial frequency shift $\delta T_f$ is determined to be $4.5 \times 10^{-3}$ or 0.0045, which is in good agreement with the typical spin-glass (SG) system values e.g., $8.0 \times 10^{-3}$, $2.2 \times 10^{-2}$, $2 \times 10^{-2}$ and $6 \times 10^{-3}$ for $U_2RhSi_3$ [38], $Ce_2AgIn_3$ [39], $La(Fe_{1-x}Mn_x)_{11.4}Si_{1.6}$ [40] and $(Eu_{1-x}Sr_x)S$ [41] respectively. Hence the studied system GdRu-1222 is a typical spin-glass system with some possibility of ferromagnetic clusters which will be discussed in next sections. There are basically two different possibility of the spin-glass (SG) freezing: first one is the existence of true equilibrium phase transition at a particular temperature resembling the canonical spin-glass [42] and the second interpretation is the existence of ferromagnetic homogenous/non-homogenous clusters embedded spin-glass (SG) matrix with non-equilibrium freezing [43]. To study the isolated clusters (superparamagnet) in the system, the frequency dependence of their freezing temperature ($T_f$) or a more appropriate in case of superparamagnets blocking temperature $T_B$, can be interpreted by *Arrhenius law* [23, 24],

$$\omega = \omega_o \, exp. \, [-E_a/k_B T_f] \ldots (2)$$

where $E_a$ is the barrier energy which separates two nearest clusters in the matrix, $\omega_o$ is the characteristics frequency of a isolated cluster, $T_f$ is the freezing temperature or blocking temperature. Actually in the studied system the freezing and blocking temperature are at the same temperature. Similarly, to study the spin-glass (SG) phase with non-interacting ferromagnetic homogenous/non-homogenous clusters state in the studied GdRu-1222, well known *Vogel-Fulcher* law is proposed [23, 24],

$$\omega = \omega_o \, exp. \, [-E_a/k_B(T_f - T_o)] \ldots (3)$$

where $T_o$ is the *Vogel-Fulcher* temperature which describes the adjacent inter-clusters interactions and $k_B$ is the Boltzmann constant. When $T_o = 0$ means there is no inter-cluster interactions (isolated clusters or superparamagnetic state) takes place in the system and then the *Vogel-Fulcher law* transform into the *Arrhenius law*. Equation (2) fitted as a linear dependence of freezing temperature ($T_f$) with the expression $1/ln(f_o/f)$, $f_o = 1/\tau_o = \omega_o/2\pi$. In figure 7 we present various *Vogel-Fulcher* plots for different characteristics frequency ($\omega_o/2\pi$) ranging between $10^{10}$-$10^{13}$ Hz, which shows that our data follow the expected linear behavior corresponding to each characteristics frequency. The parameters, Activation energy ($E_a$) and *Vogel-Fulcher* temperature ($T_o$) be also calculated corresponding to each chosen characteristics frequency. Table II provides the value of Activation energy ($E_a$), *Vogel-Fulcher* temperature ($T_a$) and the parameter $t^* = (T_f$-$T_o)/T_f$ corresponding to each characteristics frequency ranging from $10^{10}$-$10^{13}$ Hz. The values of *Vogel-Fulcher* temperature $T_o$ = 91.44 K, 90.93 K, 90.87 K, and 90.65 K corresponding to each characteristics frequency of $10^{10}$, $10^{11}$, $10^{12}$ and $10^{13}$ Hz respectively, are in good agreement of freezing point temperature $T_f$ calculated from the DC magnetization curve. It is also observed for a spin-glass (SG) system that the parameter $t^* = (T_f$-$T_o)/T_f$ must be smaller than 0.10 and above 0.15 for cluster spin-glass materials [23, 44]. It is clear that the fitted experimental data of *Vogel-Fulcher* law and the parameters $t^*$ (0.15 to 0.23) confirm the spin-glass (SG) state with FM clusters in the studied GdRu-1222 system.

To investigate further this spin-glass (SG) state possibly co-existing with the ferromagnetic clusters, we have performed non-linear AC susceptibility studies on our GdRu-1222 sample. Figure 8(a) and 8(b) shows the real ($\chi'_1$) and imaginary ($\chi''_1$) parts of first harmonic of AC susceptibility respectively, measured as a function of temperature in the range of 200 K to 2 K with zero external DC bias. Both parts (real and imaginary) are measured with drive AC field amplitude from 1 to 17 Oe and at a fix frequency of 333 Hz. It is observed that before the main PM to FM transition or says spin-glass (SG) transition there is a hump seen at around 125 K, called Neel temperature ($T_N$), which is consistent with earlier report on this system [37]. A peak is observed in both real and imaginary parts of AC susceptibility at temperature range 90-95 K, having a transition width of 20 K. Inset of figure 8(a) and 8(b) shows that peak temperature corresponding to ($\chi'_1$) and ($\chi''_1$), which shifts towards lower temperature along with increase in peak height with increasing amplitude of the AC drive field. This is unusual and not acceptable for a typical spin-glass (SG) system. For a typical spin-glass (SG) the height of the peak decreases with increasing the AC amplitude [22]. It is clear that freezing of dipole moments do not take place in the direction of the applied field, hence the magnitude of the peak decreases with increasing amplitude of the applied AC field. It is clear that instead of the pure spin-glass state, there are some non-interacting ferromagnetic (FM) clusters existing in studied GdRu-1222 below freezing temperature. The FM clusters being coexisting with spin-glass (SG) state may or may not be interacting with each other. To ascertain this dynamic magnetic phenomenon like superparamagnetism (SPM) need to be investigated. Hence to confirm the co-existence of superparamagnetism (SPM) state in the studied GdRu-1222 system, we applied Wohlfarth's model (WM) on our AC susceptibility data. It has been shown [45] that the existence of the superparamagnetic particles can be verified through the $T^{-3}$ dependence of $\chi'_3$. According to the Wohlfarth's superparamagnetic blocking model [46], $\chi'_1$ of the assembly of superparamagnetic particles follows a Curie law above the blocking temperature $T_B$, while $\chi'_3$ exhibits negative $T^{-3}$ dependence and become independent of temperature below the blocking state. At above the blocking temperature i.e., $T \geq T_B$ [47],

$$\chi'_1 = \frac{n \langle \mu \rangle}{3} \frac{\langle \mu \rangle}{k_B T} = \frac{P_1}{T} \; or \; \chi'_1 \propto \frac{1}{T} \ldots \ldots (5)$$

$$\chi'_3 = -\frac{n \langle \mu \rangle}{45} \left(\frac{\langle \mu \rangle}{k_B T}\right)^3 = \frac{P_3}{T^3} \; or \; \chi'_3 \propto \frac{1}{T^3} \ldots \ldots (6)$$

where $n$ is the number of particles per unit volume, $\langle \mu \rangle$ is the average magnetic moment of the single particle, $k_B$ is the Boltzmann constant. $P_1$ and $P_3$ are two temperature-dependent constants for the system. Figure 9(a) and 9(b) shows the appropriate plots for first-order ($\chi'_1$) and third-order harmonic ($\chi'_3$) of AC susceptibility above the blocking temperature respectively. Also the solid lines are best fit of equations (5) and (6) to the experimental data well above the blocking temperature ($T_B$). The linear dependence of $\chi'_3$ on $T^{-3}$ is found only in a small temperature interval: between 91 K and 97 K. It is reported that for conventional superparamagnetic systems the particle's internal spin-spin correlation temperature is much higher than the blocking temperature $T_B$ [48, 49]. From equation (5) and (6) we can extract the average magnetic moment of the superparamagnetic particle [48]. However, due to the presence of the ordering at $T_N$ and the paramagnetic contribution from the Gd ions, it is not possible to extract the exact magnetic moment. It has been shown recently [47] that for $Li_{0.5}Ni_{0.5}O$ a similar behavior occur within 10 K wide temperature interval, where the third harmonic is linear in $T^{-3}$. Also, for superparamagnetic clusters assembly of magnetite $Fe_3O_4$, the Curie temperature of bulk magnetite is 850 K, the blocking temperature is observed only an around 20 K [49]. In principle on can measure the superparamegnetic state above $T_B$, covering a large temperature range from blocking temperature $T_B$ to Curie temperature $T_C$. But for our GdRu-1222 system the blocking or Curie temperature and spin freezing temperature are very near to each other hence fitting is done for a small temperature interval (91 K to 97 K). The fitting range is probably too narrow but it can still serve as an indication of the existence of superparamagnetic state. Figure 10 depicts a plot between $1/\chi_{ac}$ vs. $T$ for studied GdRu-1222 system, which clearly shows two distinct slopes. It means the blocking temperature is very near to the spin-correlation temperature. It concludes that a superparamagnetic state is developed over a narrow temperature range $T_C \geq T \geq T_f$. It concludes that superparamagnetic (SPM) state co-exists with the spin-glass (SG) state in studied GdRu-1222 system in a narrow temperature range of 91 K to 97 K. This does necessary mean that the FM clusters being embedded in main spin-glass (SG) matrix are non interactive during the SPM region of 97 K to 91 K. Below 91 K the FM clusters are interacting with each other and hence the hysteresis is seen in *M(H)* at further lower temperatures. This is interesting that the spin-glass (SG) state passes through a SPM region before turning to co-existing interacting FM clusters with in main spin-glass SG state.

## IV. Conclusions

A systematic detailed results and analysis of structural, DC/linear and non-linear AC magnetization, isothermal and thermoremanat magnetization of studied complex magneto-superconductor GdRu-1222 is presented. The GdRu-1222 has a rich variety of magnetic phenomena. A paramagnetic (PM) to ferromagnetic (FM) transition at around $T_C = 110$ K, spin-glass (SG) transition temperature $T_f = 92.9$ K, with non-interacting homogenous/non-homogenous ferromagnetic clusters, the presence of superparamagnetic state just above the spin-glass (SG) formation temperature and finally a superconducting transition at around 27 K. The frequency-dependent peak observed in the temperature dependence of the AC susceptibility $\chi_{ac}$, combined with magnetic relaxation data provides strong evidence of the important role of magnetic frustration in polycrystalline Ru-1222. This also established the existence of spin-glass (SG) with magnetic clusters over a significant temperature range. Analysis of First and third harmonic of AC susceptibility within Wohlfarth's model (WM) of superparamagnetism over a wide temperature range below $T_C$ indicate that FM clusters likely to be in the superparamagnetic state. In last, our results support the presence of spin-glass (SG) state with non-homogeneous ferromagnetic clusters followed by SPM state in GdRu-1222 system. Possible random distribution of $Ru^{5+}$-$Ru^{5+}$, $Ru^{4+}$-$Ru^{5+}$ and $Ru^{4+}$-$Ru^{4+}$ exchange interactions may be responsible for observed spin-glass (SG) with ferromagnetic clusters (FM) followed by superparamagnetism (SPM) complex magnetic state.

## Acknowledgements

The authors from NPL would like to thank Prof. R. C. Budhani (Director, NPL) for his keen interest in the present work. One of us, Anuj Kumar would also thank Council of Scientific & Industrial Research (CSIR), New Delhi for financial support through Senior Research Fellowship (SRF). This work is also financially supported by DST-SERC (Department of Science and Technology), New Delhi funded project on *Study of Rutheno-cuprates in bulk and thin film form*.

# Figure Captions

**Figure 1** Observed (*solids circles*) and calculated (*solid lines*) XRD patterns of RuSr$_2$Gd$_{1.4}$Ce$_{0.6}$Cu$_2$O$_{10-\delta}$ compound at room temperature. *Solid lines* at the bottom are the difference between the observed and calculated patterns. *Vertical lines* at the bottom show the position of allowed Bragg peaks.

**Figure 2** *ZFC* and *FC DC* magnetization plots for RuSr$_2$Gd$_{1.4}$Ce$_{0.6}$Cu$_2$O$_{10-\delta}$, measured in the applied magnetic field, $H$ = 20 Oe. Inset shows the $R$ vs. $T$ plot for GdRu-1222 in zero fields.

**Figure 3** Magnetization as a function of applied magnetic field measured at different temperatures (5, 20, 50, 75, 100, 125, 150 and 200 K) in the range - 50 kOe to + 50 kOe. Inset shows the *M(H)* plots in low field range (up to 2 kOe) at 5, 20 and 50 K.

**Figure 4** Arrott plots (*H/M* vs. *M$^2$*) using *DC* magnetization data observed at different fixed temperatures (5, 20, 50, 75, 100 and 125 K).

**Figure 5** Thermoremanent magnetization (TRM) relaxation for T = 60 K and for waiting time $t_w$ = 100 s and 1000 s.

**Figure 6(a)** Temperature dependence of the real part of *AC* susceptibility, measured at different frequency with zero external *DC* magnetic fields. Inset shows the enlarged view of the real part of the first harmonic *AC* susceptibility.

**Figure 6(b)** Temperature dependence of the imaginary part of *AC* susceptibility, measured at different frequency with zero external *DC* magnetic fields. Inset shows the enlarged view of the imaginary part of the first harmonic *AC* susceptibility.

**Figure 7** The variation of the freezing temperature $T_f$ with the frequency of the *AC* field, at different characteristics frequencies ($10^{10}$-$10^{13}$ Hz), in a *Vogel-Fulcher* plot. The solid lines are the best fit of equation.

**Figure 8(a)** Temperature dependence of the real part of *AC* susceptibility measured at different amplitude with zero external *DC* magnetic fields. Inset shows the enlarged view of the real part of the first harmonic *AC* susceptibility.

**Figure 8(b)** Temperature dependence of the imaginary part of *AC* susceptibility, measured at different amplitude with zero external *DC* magnetic fields. Inset shows the enlarged view of the imaginary part of the first harmonic *AC* susceptibility.

**Figure 9(a)** First order harmonics of *AC* susceptibility is fitted to Wohlfarth's model (WM) above the freezing temperature (T$_f$) for studied GdRu-1222. The solid line shows $T^{-1}$ fit to $\chi'_1$.

**Figure 9(b)** Third order harmonics of *AC* susceptibility is fitted to Wohlfarth's model (WM) above the freezing temperature (T$_f$) for studied GdRu-1222. The solid line shows $T^{-3}$ fit to $\chi'_3$.

**Figure 10** The inverse of first-order *AC* susceptibility ($\chi$) plotted with temperature indicating two distinct slopes corresponding to paramagnetic and superparamagnetic phase.

**Table I.** Atomic coordinates and site occupancy for studied RuSr$_2$Gd$_{1.4}$Ce$_{0.6}$Cu$_2$O$_{10-\delta}$
Space group: ***I4/mmm***, Lattice parameters; *a* = 3.8350 (4) Å, *c* = 28.5719 (6) Å, $\chi^2$ = 2.19

| Atom | Site | *x* | *y* | *z* |
|---|---|---|---|---|
| **Ru** | *2b* | 0.0000 | 0.0000 | 0.0000 |
| **Sr** | *2h* | 0.0000 | 0.0000 | 0.4199 (4) |
| **Gd/Ce** | *1c* | 0.0000 | 0.0000 | 0.2931 (7) |
| **Cu** | *4e* | 0.0000 | 0.0000 | 0.1420 (3) |
| **O (1)** | *8j* | 0.6152 (3) | 0.5000 | 0.0000 |
| **O (2)** | *4e* | 0.0000 | 0.0000 | 0.0659 (2) |
| **O (3)** | *8g* | 0.0000 | 0.5000 | 0.1419 (3) |
| **O (4)** | *4d* | 0.0000 | 0.5000 | 0.2500 |

**Table II.** Activation energy $E_a$ (eV), *Vogel-Fulcher* temperature $T_o$ (K) and parameter $t^* = (T_f - T_o)/T_f$

| Characteristic frequency | Activation energy $E_a$ (eV) | *Vogel-Fulcher* temperature $T_o$ (K) | Parameter $t^* = (T_f - T_o)T_f$ |
|---|---|---|---|
| $f_o = 10^{10}$ Hz | 2.48x10$^{-3}$ eV | 91.44 K | 0.015 |
| $f_o = 10^{11}$ Hz | 3.62x10$^{-3}$ eV | 90.93 K | 0.021 |
| $f_o = 10^{12}$ Hz | 4.18x10$^{-3}$ eV | 90.87 K | 0.022 |
| $f_o = 10^{13}$ Hz | 5.08x10$^{-3}$ eV | 90.65 K | 0.024 |

**Figure 1**

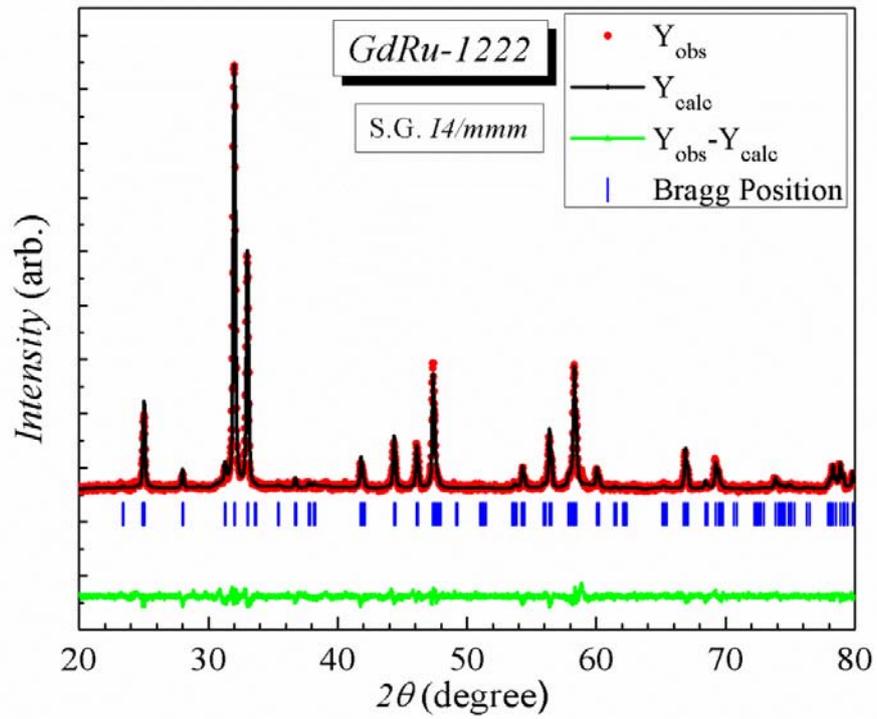

**Figure 2**

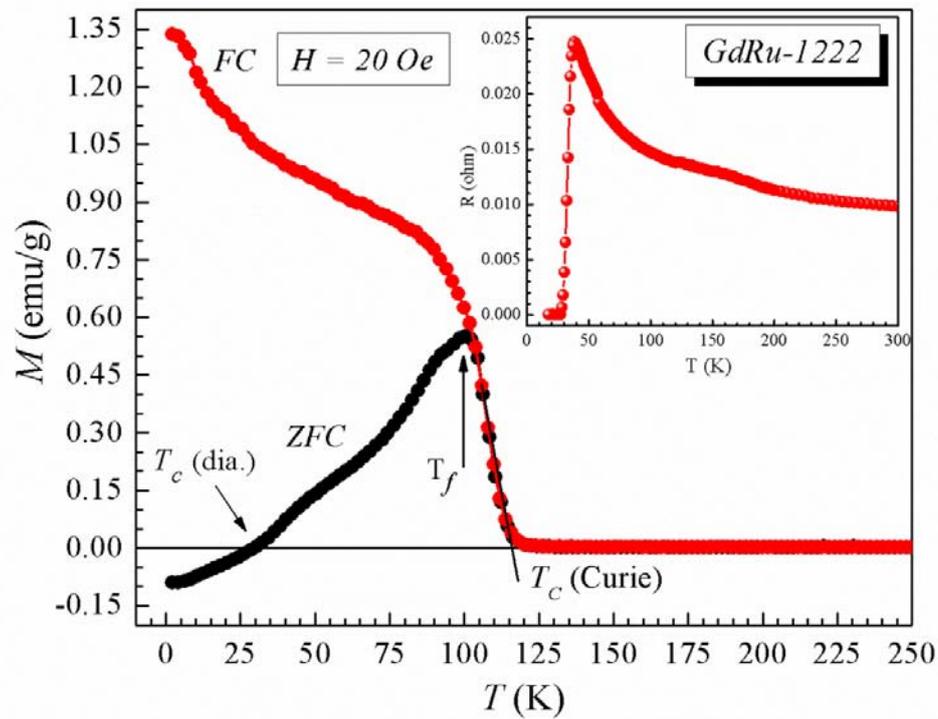

**Figure 3**

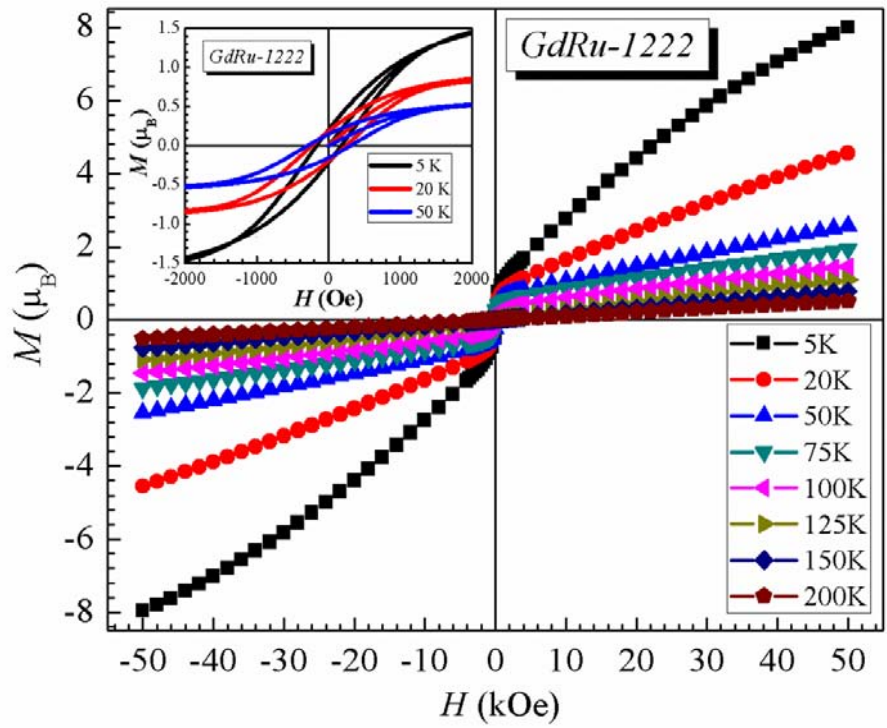

**Figure 4**

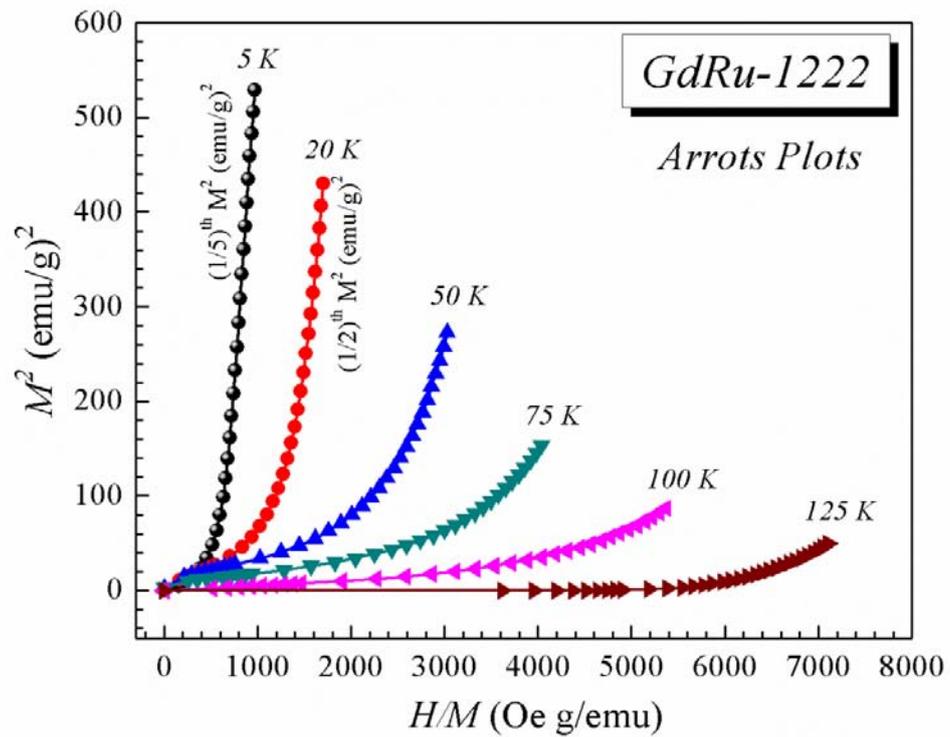

**Figure 5**

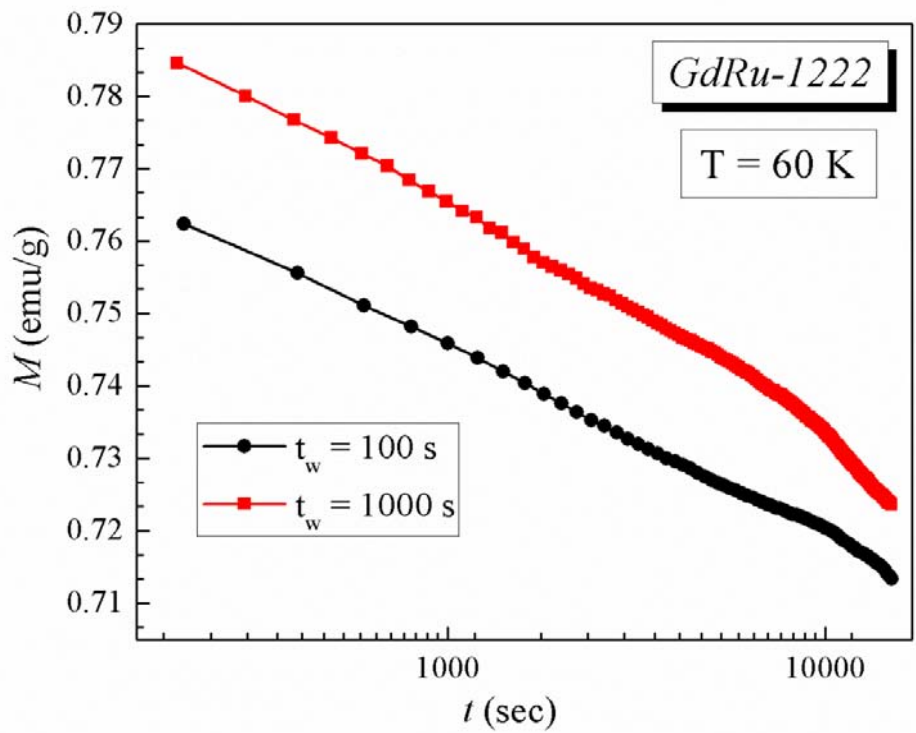

**Figure 6(a)**

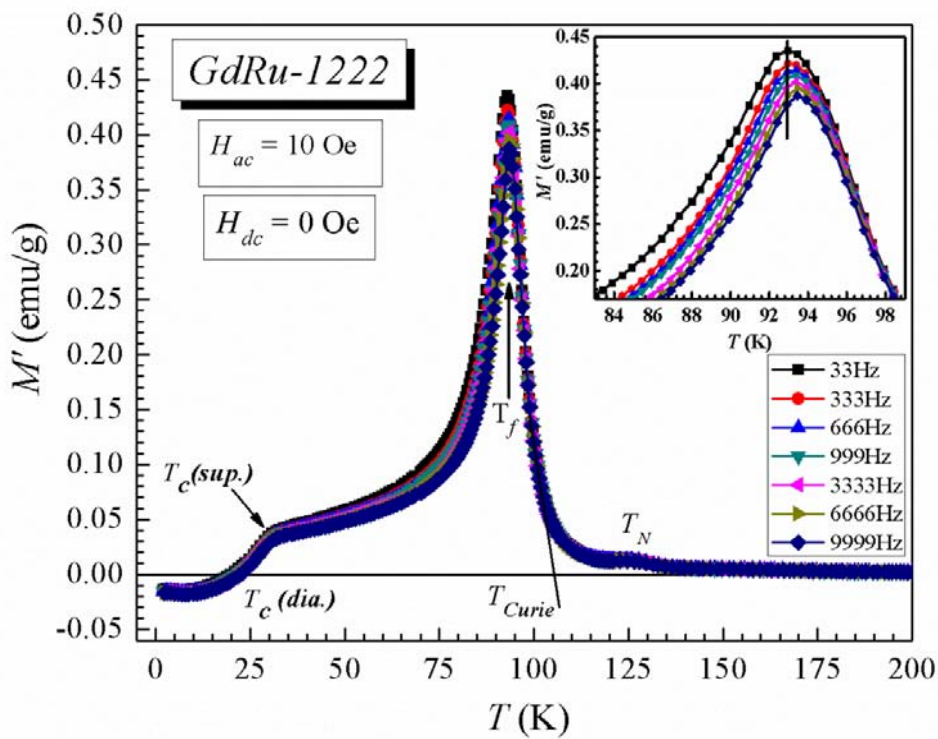

**Figure 6(b)**

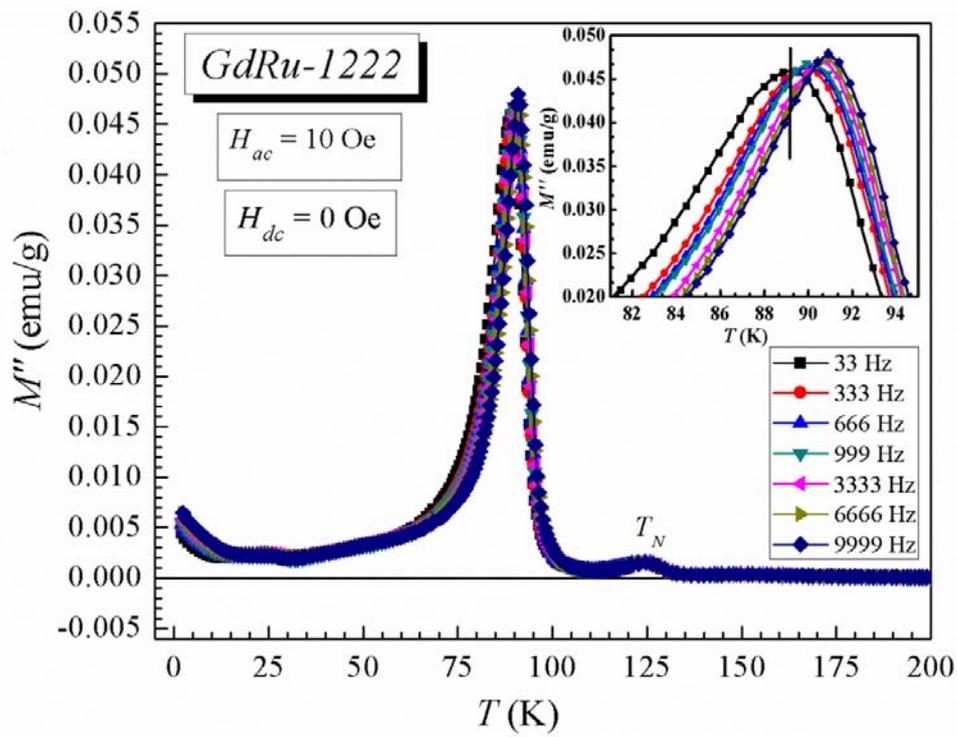

**Figure 7**

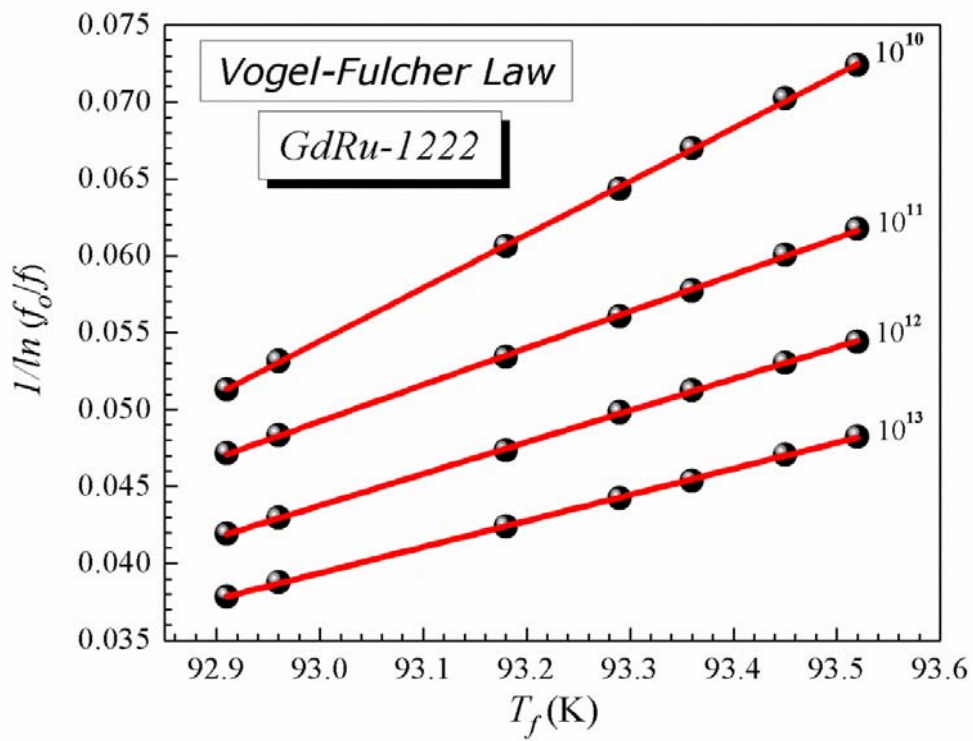

**Figure 8(a)**

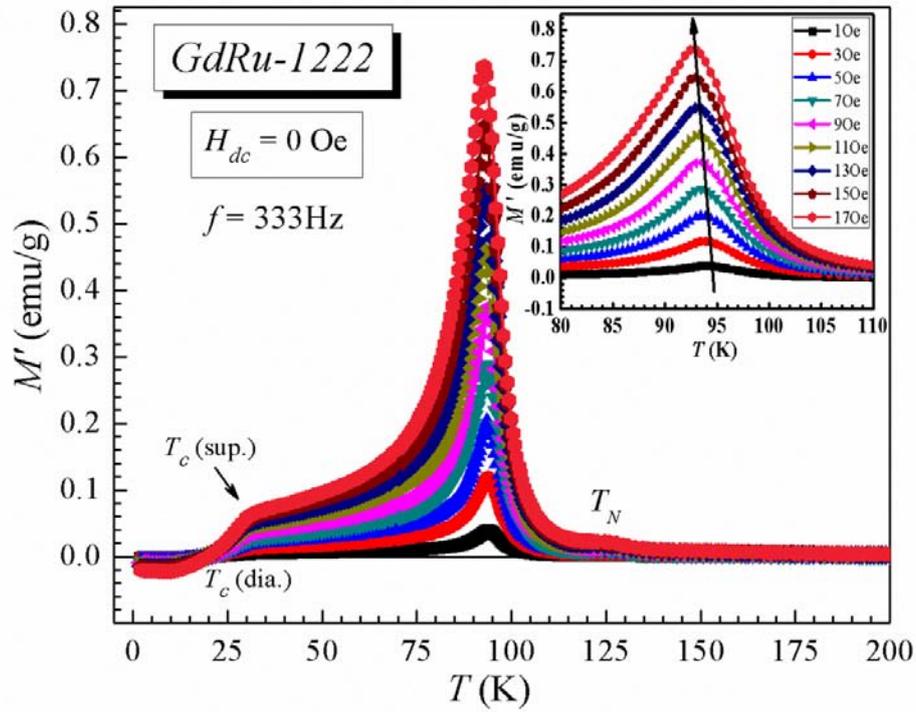

**Figure 8(b)**

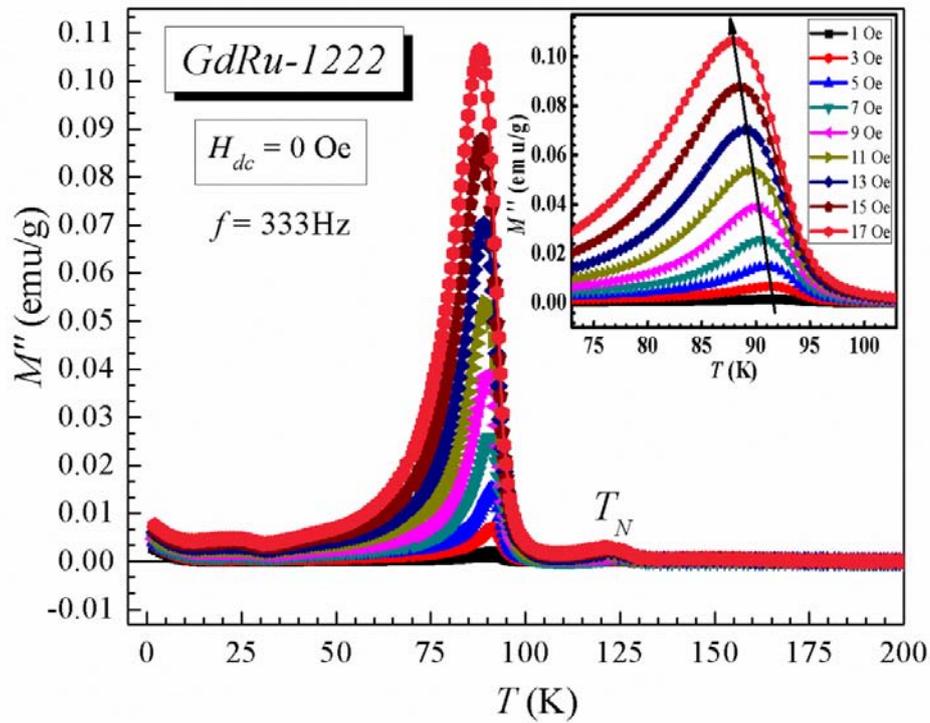

**Figure 9(a)**

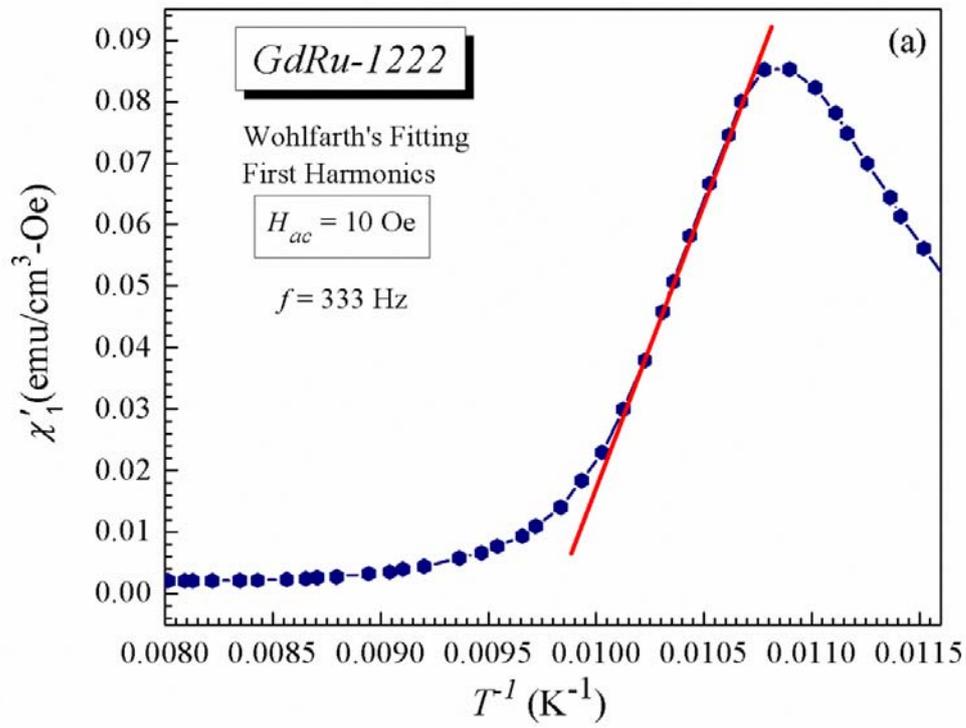

**Figure 9(b)**

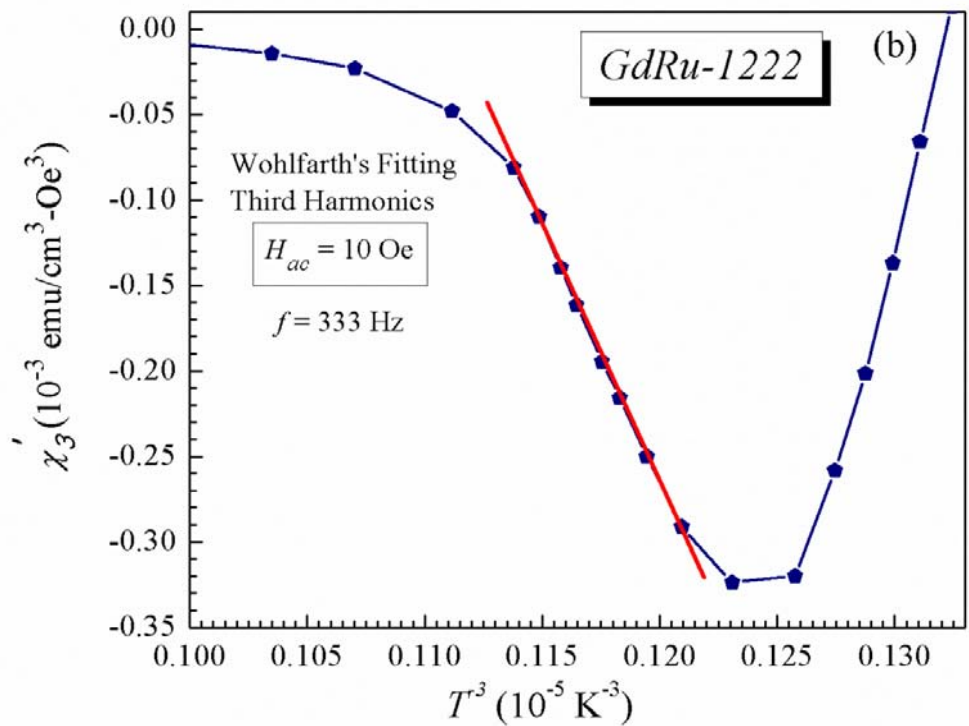

**Figure 10**

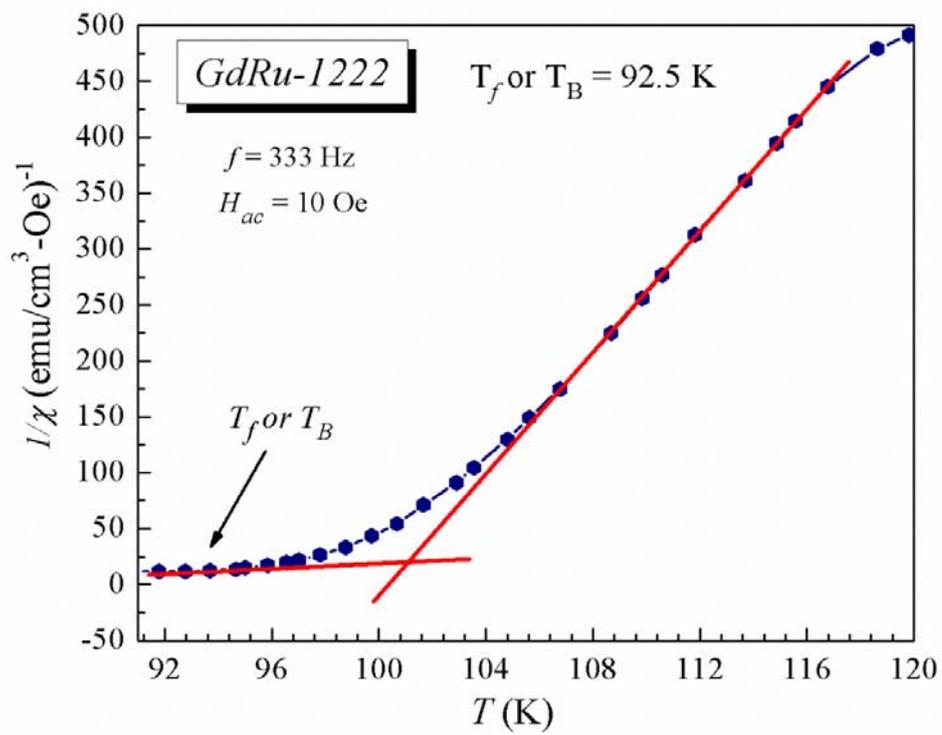